\documentclass[a4paper,11pt]{article}
\usepackage{jheppub} 
\usepackage{lineno}

\usepackage{graphicx} 
\usepackage{subcaption}
\usepackage{amsmath}
\usepackage{amssymb}
\usepackage{dsfont}
\usepackage{mathtools}
\usepackage{float}

\makeatletter
\def\@fpheader{}
\makeatother

\title{\boldmath Thermal Bootstrap of Large-$N$ Matrix Models via Conic Optimization}

\author{Sophia M. Adams}
\affiliation{Division of Physics, Mathematics and Astronomy, Caltech,\\
Pasadena, California, USA}

\emailAdd{sadams2@caltech.edu}

\abstract{This paper is aimed at improving thermal bootstrapping methods for matrix quantum mechanics. The thermal energies of the large-$N$ one-matrix anharmonic oscillator and large-$N$ two-matrix anharmonic oscillator were bounded without logarithmic relaxation using the Quantum Information Conic Solver. For the one-matrix model, which can be interpreted using an effective theory of ``long strings'' in the low temperature limit, stricter bootstrap bounds yield a value of the first long string excited energy within $0.001\%$ of the physical value and the first estimation from symmetry and self-consistency equations alone of the first long string coupling coefficient.}

\begin{document}
\maketitle
\flushbottom

\section{Background}
Bootstrapping is a method of solving a theory by specifying conditions that the theory must satisfy, such as self-consistency equations and symmetries. Solving a theory with bootstrapping can be advantageous because it can produce tight, rigorous bounds on observables without much computational cost. In particular, matrix quantum mechanics (MQM) is an important bootstrapping target because some MQM theories may not be analytically solvable or reliably solvable through numerical or perturbative methods but may contain many symmetries. Additionally, MQM provides simplified, nonperturbative models that capture essential features of quantum gravity and black hole dynamics, such as black hole entropy in the BFSS matrix model (Ref.~\cite{PhysRevD.55.5112}).

In the thermal bootstrap, the main object of study is the thermal density matrix, constrained by symmetries and by the Kubo–Martin–Schwinger (KMS) condition that characterizes thermal equilibrium (Refs.~\cite{ArakiSewell1977}, \cite{Sewell1977}). Bootstrapping a thermal state involves determining which combinations of correlators and expectation values are consistent with these constraints. In this work, the thermal energy is bootstrapped for the one-matrix and two-matrix anharmonic oscillator. From the one-matrix model, I estimate the long string ground state energy, the first long string excited state energy, and the first long string coupling coefficient, which characterize the structure of the thermal spectrum and the interaction strength in the matrix model's low temperatrure effective long string description.

Previous thermal bootstrap studies have primarily used linear semidefinite programming (SDP) methods, which replace nonlinear constraints with global linear relaxations. While rigorous, such methods can fail to enforce nonlinear relations like the KMS condition exactly. To overcome this limitation, I use the Quantum Information Conic Solver (QICS) (Ref.~\cite{he2024qicsquantuminformationconic}), a nonlinear SDP solver that handles conic constraints directly. 

In this work, I take steps toward improving thermal bootstrapping methods for MQM by applying QICS to bootstrap the one-matrix and two-matrix anharmonic oscillators, achieving larger system sizes and tighter bounds than previously obtained through linear methods (Ref.~\cite{cho2025thermalbootstrapmatrixquantum}).

\section{Models}
In this paper, I am interested in the one-matrix and two-matrix anharmonic oscillator in the 't Hooft large-$N$ limit. For each bootstrap problem, I consider operators built as products of elementary matrices $X, P$ (and other matrices in the two-matrix model). The length $L$ is the maximum number of operators appearing in a single product. Increasing $L$ allows longer operator products in the bootstrap and generally improves the precision of the resulting bounds. All constraints described in this section with the exception of the KMS constraint and positive semidefinite constraint are imposed at the level of SDP assembly. The SDP solver is used only to impose conic constraints--the positive semidefinite constraint and the KMS constraint.

\subsection{One-Matrix}
In one-matrix quantum mechanics, the quartic anharmonic oscillator Hamiltonian takes the form 
\begin{equation}
H=\operatorname{Tr}(\frac{1}{2}P^2+\frac{1}{2}X^2+\frac{g}{N}X^4)
\end{equation}
where the canonical coordinate $X$ and its conjugate momentum $P$ are traceless $N \times N$ Hermitian matrices. The gauged theory or ``singlet sector'' can be reformulated in terms of non-interacting fermions, while the ungauged theory contains ``non-singlet sectors'' whose low-lying states in the adjoint sector admit an effective description in terms of ``long strings'' (Refs.~\cite{Maldacena2005LongStrings} and \cite{Balthazar_2019}) and is not known to be analytically solvable. One can derive bounds on the energy spectrum, from which one can extract information about the energy spectrum of the effective long strings and long string interactions in the adjoint sector, by bootstrapping the ungauged theory at finite temperature where 
\begin{equation}
E(\beta) \equiv \langle H \rangle_\beta \equiv \operatorname{Tr}_{\mathcal{H}}(\rho_\beta H)=\operatorname{Tr}_{\mathcal{H}}(\frac{1}{Z(\beta)} e^{-\beta H} H) = \operatorname{Tr}_{\mathcal{H}}(\frac{1}{\operatorname{Tr}_{\mathcal{H}}\left(e^{-\beta H}\right)} e^{-\beta H} H).
\end{equation}
Due to large-$N$ factorization, 
\begin{equation}
\langle \operatorname{Tr}(\mathcal{O}_1) ... \operatorname{Tr}(\mathcal{O}_1) \rangle_\beta = \langle \operatorname{Tr}(\mathcal{O}_1) \rangle_\beta ... \langle \operatorname{Tr}(\mathcal{O}_1) \rangle_\beta ,
\end{equation}
so we only need to consider single-trace operators. Note that each trace is to be regarded as order $N$, and $X$ and $P$ are rescaled by $\frac{1}{\sqrt{N}}$, allowing us to omit the $\frac{1}{N}$ multiplying $X^4$ in the Hamiltonian. As in Ref.~\cite{cho2025thermalbootstrapmatrixquantum}, I express the components of $X$ and $P$ as $X_{ab}=X^AT_{ab}^A$ and $P_{ab}=P^AT_{ab}^A$ where $T_{ab}^A$ are the traceless Hermitian $SU(N)$ generators satisfying the completeless relation $T_{ab}^AT_{cd}^A=\delta_{ad}\delta_{bc}-\frac{1}{N}\delta_{ab}\delta_{cd}$ so that the canonical commutation relation $\frac{1}{N}[X^A,P^B]=\frac{i}{N}\delta^{AB}$ is equivalent to
\begin{equation}
    \frac{1}{N}[X_{ab},P_{cd}]=\frac{i}{N}\delta_{ad}\delta_{bc} + \mathcal{O}(\frac{1}{N^2}).
\label{largeNcommutation}
\end{equation}

I consider the same large-$N$ simplifications as in Ref.~\cite{cho2025thermalbootstrapmatrixquantum}, including treating any remaining trace products as independent variables. I find bounds on the thermal energy by bootstrapping the system of operators $\mathcal{O}_i$ in tuples of $X$ and $P$ up to maximum length $\frac{L}{2}$ (e.g. for $L=6$, $O_i \in \{1,X,P,XP,PX,XXX,XXP,XPX,PXX,XPP,PXP,PPX,$ $PPP\}$) subject to the following constraints:

\begin{itemize}
    \item Semidefinite positivity of the moment matrix $\mathcal{M}_{ij} = \langle \operatorname{Tr}(\mathcal{O}_i^\dagger \mathcal{O}_j) \rangle_\beta$
    \item Parity symmetry, $\langle \operatorname{Tr}(\mathcal{O'}_1 \mathcal{O'}_2\cdots\mathcal{O'}_n) \rangle_\beta=0$ with $n$ odd and $\mathcal{O'}_i \in \{X,P\}$ 
    \item Time-reversal symmetry, $\operatorname{Re}(\langle \operatorname{Tr}(P_1 P_2\cdots P_n) \rangle_\beta)=\operatorname{Im}(\langle \operatorname{Tr}(P_1 P_2\cdots P_m) \rangle_\beta)=0$ with $n$ odd and $m$ even
    \item Reality conditions, $\langle \operatorname{Tr}(\mathcal{O}_i^\dagger) \rangle_\beta = \langle \operatorname{Tr} (\mathcal{O}_i) \rangle_\beta ^*$
    \item Stationary state conditions, $\langle [H,\operatorname{Tr} (\mathcal{O}_i)]\rangle_\beta=0$
    \item Cyclicity of the trace with large-$N$ factorization, 
    $$\langle\operatorname{Tr}(\mathcal{O'}_1 \mathcal{O'}_2 \cdots \mathcal{O'}_n)\rangle_\beta=\langle\operatorname{Tr}(\mathcal{O'}_2 \mathcal{O'}_3 \cdots \mathcal{O'}_n \mathcal{O'}_1)\rangle_\beta$$ $$ + \sum_{k=2}^n\langle\operatorname{Tr}(\mathcal{O'}_2 \cdots \mathcal{O'}_{k-1})\rangle_\beta \langle\operatorname{Tr}([\mathcal{O'}_1,\mathcal{O'}_k])\rangle_\beta \langle \operatorname{Tr}(\mathcal{O'}_{k+1} \cdots \mathcal{O'}_{n})\rangle_\beta$$ with $\mathcal{O'}_i \in \{X,P\}$  
    \item Normalization, $\frac{1}{N}\langle \operatorname{Tr}(\mathds{1}) \rangle_\beta = 1$
    \item Large-$N$ commutation relations (Eq.~\eqref{largeNcommutation}) 
    \item KMS condition on a thermal state (Section~\ref{thermalConstraints})
\end{itemize}

\subsection{Two-Matrix}
\label{modelstwomat}
 The two-matrix quartic anharmonic oscillator is described by the Hamiltonian 
 \begin{equation}
 H=\operatorname{Tr}(\frac{1}{2}(P_1^2+P_2^2)+\frac{1}{2}(X_1^2+X_2^2)-\frac{g}{N}[X_1,X_2]^2) 
 \end{equation}
 where $P_1$ is the conjugate momentum of $X_1$ and $P_2$, $X_2$. Bootstrapping the two-matrix model is ultimately motivated by the BFSS matrix model, conjectured to be dual to M-theory in the large-$N$ limit \cite{PhysRevD.55.5112}. Due to the computational cost associated with bootstrapping a large number of matrices and the fact that the thermal state is not supersymmetric, bootstrapping the full BFSS model, which has 9 bosonic and 16 fermionic matrices, is a much more nontrivial task using this method. Instead, I consider the bosonic BFSS model, dual to a theory of a black D0-brane in the large-$N$ limit, truncated to two matrices. As an additional simplification, I include the mass term $\frac{1}{2}(X_1^2+X_2^2)$.

 The two-matrix system made from tuples of $X_1$, $X_2$, $P_1$, and $P_2$ is subject to the same constraints as the one-matrix model with an additional constraint that comes from swapping the $1$ and $2$ operator labels: $X_1 \leftrightarrow X_2$, $P_1 \leftrightarrow P_2$. Reformulating the model in terms of complex operators 
 \begin{equation}
 P=\frac{1}{\sqrt{2}} (P_1+iP_2), \quad \bar P = \frac{1}{\sqrt{2}} (P_1-iP_2), \quad X=\frac{1}{\sqrt{2}} (X_1+iX_2), \quad \bar X = \frac{1}{\sqrt{2}} (X_1-iX_2)
 \label{complexops}
 \end{equation}
 as in Ref. \cite{lin2025highprecisionbootstrapmultimatrixquantum} allows us to view the $SO(2)$ symmetry as a $U(1)$ symmetry and make manifest additional constraints from $U(1)$ charge conservation. In terms of complex operators, the Hamiltonian can be written 
 \begin{equation}
 H=\operatorname{Tr}(P \bar P +X \bar X + \frac{g}{N}[X,\bar X]^2) 
 \end{equation}
 since the system is invariant under the relabeling of barred and unbarred operators. 
 
 In practice, the $U(1)$ charge of a term is the difference between the number of barred and unbarred operators, and charge conservation implies that any observable with a nonzero $U(1)$ charge is zero. In this basis, we can consider the same constraints as in the real basis but can replace the parity constraint with the stronger $U(1)$ charge constraint. Additionally, the time-reversal and operator relabeling constraints can be reinterpreted in the new basis as follows. I will refer to the original time-reversal transformation, 
 \begin{equation}
P_1 \rightarrow -P_1,\quad 
P_2 \rightarrow -P_2,\quad 
i \rightarrow -i,
\end{equation}
 as $T$ and the new time-reversal-like transformation, 
 \begin{equation}
 P \rightarrow -P, \quad \bar P \rightarrow - \bar P, \quad i\rightarrow -i,
 \end{equation}
 as $T'$. Just as $T$ implied constraints on the real and imaginary parts of expectation values of $X_1,X_2,P_1,P_2$ operators, $T'$ now implies these constraints on the expectation values of the complex operators. Namely, any observable with an even number of $P$ and $\bar P$'s is purely real, while any observable with an odd number of $P$ and $\bar P$'s is purely imaginary. This constraint significantly reduces the number of variables. However, the original $T$ constraint reinterpreted in the new basis is still needed in order to obtain meaningful bounds on the energy. In terms of the complex operators, $T$ becomes the transformation 
 \begin{equation}
 X \rightarrow \bar X, \quad \bar X \rightarrow X, \quad P \rightarrow - \bar P, \quad \bar P \rightarrow -P, \quad i \rightarrow -i. 
 \end{equation}
 The relabeling constraint becomes a relabeling of barred and unbarred operators; however, in practice, I leave this constraint out since it neither improves the bounds nor reduces the number of free variables. In fact, $T'$ is a composition of this relabeling and $T$. The original relabeling constraint turns out to be equivalent to the new relabeling constraint when $U(1)$ charge conservation is imposed.

\section{Thermal Constraints}
\label{thermalConstraints}
At finite temperature, the bootstrap problem can be formulated as a conic optimization problem where constraints take the form 
\begin{equation}
\mathcal{M} \succeq 0,
\end{equation}
representing a semidefinite cone and 
\begin{equation}
\beta C \succeq A^{1/2}\log(A^{1/2}B^{-1}A^{1/2})A^{1/2}, 
\end{equation}
representing an operator relative entropy cone for matrices 
\begin{equation}
A_{ij}=\frac{1}{N^2} ( \langle \operatorname{Tr} (\mathcal{O}_i^\dagger \mathcal{O}_j) \rangle_\beta -\frac{1}{N} \langle \operatorname{Tr}(\mathcal{O}_i^\dagger) \rangle_\beta \langle \operatorname{Tr}(\mathcal{O}_j) \rangle_\beta),
\end{equation}
\begin{equation}
B_{ij}=\frac{1}{N^2} ( \langle \operatorname{Tr} (\mathcal{O}_j \mathcal{O}_i^\dagger) \rangle_\beta -\frac{1}{N} \langle \operatorname{Tr}(\mathcal{O}_j) \rangle_\beta \langle \operatorname{Tr}(\mathcal{O}_i^\dagger) \rangle_\beta),
\end{equation}
and 
\begin{equation}
C=\frac{1}{N^2} \langle \operatorname{Tr} (\mathcal{O}_i^\dagger [H, \mathcal{O}_j]) \rangle_\beta
\end{equation}
constructed from products of tuples of maximum length $\frac{L-4}{2}$ (Ref. \cite{cho2025thermalbootstrapmatrixquantum}). The operator relative entropy conic condition is equivalent (Refs.~\cite{ArakiSewell1977}, \cite{Sewell1977}, and \cite{Fawzi_2024}) to the Kubo-Martin-Schwinger (KMS) condition on a thermal state and is replaced by energy constraint equations, $\langle H \mathcal{O}_i\rangle = E\langle \mathcal{O}_i \rangle$, at ground state. While the ground state energy can be readily bounded using semidefinite programming, the KMS constraint is nonlinear and so requires semidefinite relaxation to turn it into a semidefinite programming problem. This is the approach taken in Ref. \cite{cho2025thermalbootstrapmatrixquantum} and solves the thermal bootstrapping problem by relaxing the nonlinear operator relative entropy cone into a linear positive semidefinite cone by approximating the $\log$: 
\begin{equation}
\log(x) \approx r_{m,k}(x) := 2^k r_m\left(x^{1/2^k}\right),
\label{logapprox}
\end{equation}
where $m,k \in \mathbb{Z}^+$,
\begin{equation}
    r_m(x) \equiv \sum_{j=1}^{m} w_j f_{t_j}(x), \qquad
    f_t(x) \equiv \frac{x - 1}{t(x - 1) + 1},
\label{rmdef}
\end{equation} 
and $t_j \in [0,1]$ and $w_j > 0$ are the nodes and weights of the $m$-point Gauss--Radau quadrature, chosen so that
\begin{equation}
    \int_0^1 p(t)\,dt = \sum_{j=1}^{m} w_j p(t_j)
\label{nodesweights}
\end{equation}
holds for any polynomial $p(t)$ of degree $2m - 2$, with $t_1 = 0.6$. The functions $r_{m,k}(x)$ provide successively tighter rigorous upper bounds on $\log(x)$ with increasing $(m,k)$ (Ref.~\cite{cho2025thermalbootstrapmatrixquantum}). 

Alternatively, the operator relative entropy cone can be handled directly by QICS (Ref.~\cite{he2024qicsquantuminformationconic}), which implements the Skajaa–Ye algorithm (Ref.~\cite{SkajaaYe2015}) together with the stepping strategy from Hypatia (Ref.~\cite{coey2023performance}). This algorithm allows QICS to handle nonsymmetric cones using only the primal barrier, without requiring the conjugate, which is unavailable for the quantum relative entropy cone. In practice, QICS exploits the structure of the quantum relative entropy cone and its relevant slices to efficiently solve the linear systems that arise in each interior-point iteration. This approach enforces the quantum relative entropy cone through iterative linearizations at each Newton step, avoiding the need for global linear approximations like semidefinite relaxations.

\section{Results}
I compare two methods for solving the bootstrap: handling the thermal constraint exactly with QICS and using a $(m,k)=(3,3)$ approximation of the $\log$ (Eqs.~\ref{logapprox}, \ref{rmdef}, and \ref{nodesweights}) with MOSEK. I also derive stricter bounds than what was previously achieved in Ref. \cite{cho2025thermalbootstrapmatrixquantum} for the one-matrix and two-matrix bootstrap with finite mass. All computations were performed on a 10-core Intel i5-1235U processor.

\subsection{One-Matrix}

\begin{figure}[h]
    \centering
    \begin{subfigure}{0.45\textwidth}
        \includegraphics[width=\linewidth]{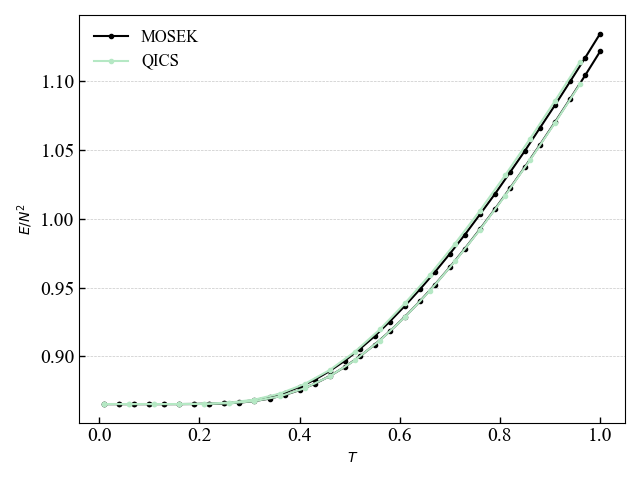}
        \caption{$L=8$}
        \label{oneMatrixg2L8}
    \end{subfigure}
    \hfill
    \begin{subfigure}{0.45\textwidth}
        \includegraphics[width=\linewidth]{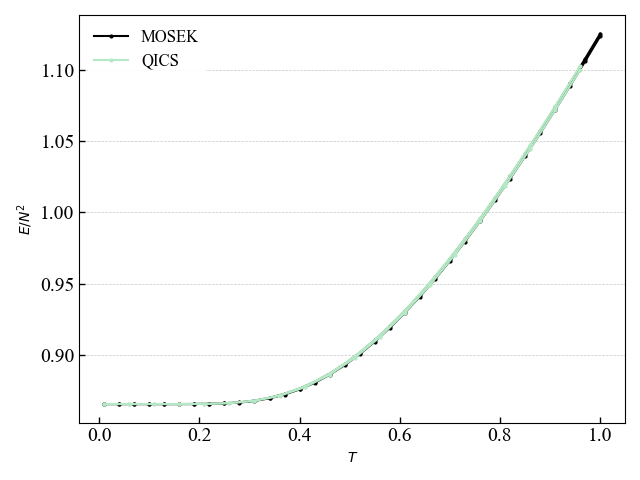}
        \caption{$L=10$}
        \label{oneMatrixg2L10}
    \end{subfigure}
    \caption{Thermal energy bounds for $g=2$ found using MOSEK with $(3,3)$ logarithmic relaxation and QICS without relaxation for system sizes $L=8,10$. The bounds agree within numerical tolerance at fixed $L$ and tighten with increasing $L$.}
    \label{oneMatrixg2L8L10}
\end{figure}

\begin{figure}[H]
\centering
\includegraphics[width=.6\textwidth]{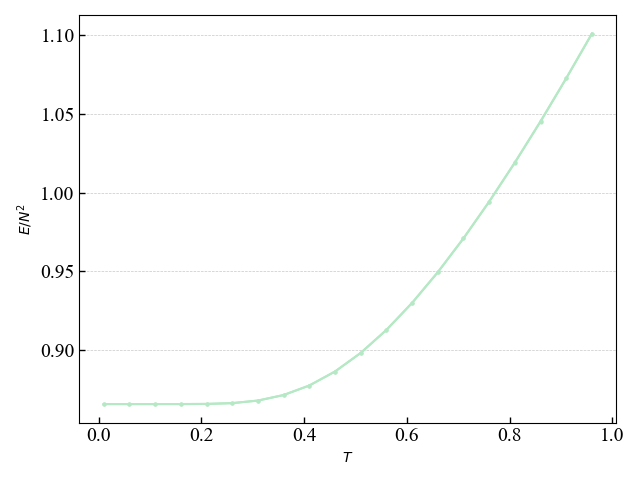}
\caption{Bounds for system size $L=12$ found using QICS are stricter than what was previously achieved with thermal relaxation in Ref. \cite{cho2025thermalbootstrapmatrixquantum}.}
\label{oneMatrixg2L12}
\end{figure}

For $L=8$, the MOSEK and QICS bounds have good agreement (Figure~\ref{oneMatrixg2L8}) and can be derived with comparable computation time (both on the order of a couple seconds or less per optimization). For $L=10$, MOSEK runs into numerical instability, failing to converge at most temperatures. Bounds can be found with MOSEK by relaxing the tolerance from $10^{-8}$ (the default for both MOSEK and QICS) to $10^{-7}$ (Figure~\ref{oneMatrixg2L10}). On the other hand, QICS returns rigorous bounds in comparable computation time (on the order of a minute per optimization) without encountering numerical instability or needing to relax the tolerance. For this reason, QICS was used to find the new bounds for $L=12$ (Figure~\ref{oneMatrixg2L12}), which took a computation time on the order of $10-10^2$ minutes per optimization.

\subsubsection{Comparison to the effective long string model}
To extract energy eigenvalues, I compare the model to its low temperature effective long string description. 

As in Ref. \cite{cho2025thermalbootstrapmatrixquantum}, I consider the semiclassically quantized phase space defined by 
\begin{equation}
J(e)=\frac{1}{\pi} \int_{\lambda_-(e)}^{\lambda_+(e)} d\lambda \sqrt{2(e-v(\lambda))}
\label{Je}
\end{equation}
where 
\begin{equation}
    v(\lambda)=\frac{\lambda^2}{2}+g\lambda^4
\end{equation}
and $\lambda_\pm (e)$ are the turning points of $v(\lambda)$. The Fermi energy $e_f$ is defined as the solution to $J(e_f)=1$, and the ground state eigenvalue distribution is given by
\begin{equation}
    \rho(\lambda)=\frac{1}{\pi} \sqrt{2(e_f -v(\lambda))}.
\end{equation}
Then in the adjoint sector at large-$N$ the energies $\Delta_n$ relative to the ground state and their eigenvectors $w_n(\lambda)$ can be found from the following singular integral eigenvalue equation
\begin{equation}
    \int_{\lambda_-(e)}^{\lambda_+(e)} d\lambda' \rho(\lambda') \frac{w_n(\lambda)-w_n(\lambda')}{(\lambda-\lambda')^2}=\Delta_n w_n(\lambda)
\label{intEigenvalue}
\end{equation}
with orthogonality $\int_{\lambda_-(e)}^{\lambda_+(e)} d\lambda \rho(\lambda) w_n(\lambda)=0$ and normalization $\int_{\lambda_-(e)}^{\lambda_+(e)} d\lambda \rho(\lambda) |w_n(\lambda)|^2=1$ conditions.

The numerical values of the energy eigenvalues were computed in Ref. \cite{Marchesini:1978ud} and are listed below for $g=2$: $$e_0 = 0.8654577$$ $$\Delta_1 = 2.1281936$$ $$\Delta_2 = 4.6201131$$ $$\Delta_3 = 7.0716122$$ $$\Delta_4 = 9.5258038.$$

As in Ref.~\cite{cho2025thermalbootstrapmatrixquantum}, the long string thermal partition function to 2-loop order in the planar limit is given by 
\begin{equation}
\frac{-\operatorname{log}Z(\beta)}{N^2} = \sum_n \operatorname{log}(1-e^{-\beta \Delta_n}) + \beta \sum_{j,k} \frac{h_{jkjk}}{(e^{\beta \Delta_j} - 1) (e^{\beta \Delta_k} - 1)} + \mathcal{O}(h^2)
\label{partitionfunc}
\end{equation}
where 
\begin{equation}
 h_{abcd}=-\int_{\lambda_-(e)}^{\lambda_+(e)}   d\lambda d\lambda' \rho(\lambda) \rho (\lambda') \frac{w_c^*(\lambda)w_d^*(\lambda')(w_a(\lambda)-w_a(\lambda'))(w_b(\lambda)-w_b(\lambda'))}{(\lambda-\lambda')^2}
\label{habcd}
\end{equation}
are coupling coefficients.

To calculate the long string coupling coefficients $h_{abcd}$, I numerically compute the eigenvectors $w_n(\lambda)$ and evaluate Eq.~\eqref{habcd} by approximating the integral as a discrete sum. 

To evaluate Eq.~\eqref{Je} numerically and solve for the Fermi energy $e_f$, I use Mathematica's NIntegrate and the closed form solution for the turning points.

I solve the integral eigenvalue equation (Eq.~\eqref{intEigenvalue}) by discretizing the integral using Gaussian quadrature and building a symmetric matrix whose eigenvalues converge to the continuum energies. (This approach can also be interpreted through the Rayleigh-Ritz/Nyström method (Ref.~\cite{Atkinson1997}).)

Eq.~\eqref{intEigenvalue} is approximated as a weighted sum over $Q$ quadrature nodes $\lambda_i$ with weights $\omega_i^{\text{quad}}$ (with both nodes and weights supplied by Gaussian quadrature) so that it becomes a discrete matrix equation:
\begin{equation}
\sum_{j=1}^{Q} 
\omega_j^{\text{quad}}\,\rho(\lambda_j) \frac{w_n(\lambda_i)-w_n(\lambda_j)}{(\lambda_i-\lambda_j)^2}
= \Delta_n\, w_n(\lambda_i).
\label{discretemateq}
\end{equation}

It is convenient to rescale the discrete eigenvectors according to
\begin{equation}
\tilde w_i := \sqrt{\omega^{\text{quad}}_i \rho(\lambda_i)}\, w_n(\lambda_i) .
\end{equation}
Multiplying Eq.~\eqref{discretemateq} by $\sqrt{\omega^{\text{quad}}_i\rho(\lambda_i)}$ yields the
symmetric eigenvalue problem
\begin{equation}
\sum_{j=1}^Q
M_{ij}\, \tilde w_j
=
\Delta_n\, \tilde w_i ,
\qquad
M_{ij} =
\begin{cases}
\displaystyle
\frac{\sqrt{\omega^{\text{quad}}_i \omega^{\text{quad}}_j \rho(\lambda_i) \rho(\lambda_j)}}{(\lambda_i - \lambda_j)^2}, & i\neq j,\\[2mm]
0, & i=j
\end{cases}
\label{finaleigenvalueequation}
\end{equation}
where the diagonal terms $M_{ii}$ are set to zero to regularize the singularity at $\lambda_i = \lambda_j$. 
In the continuum, the numerator $(w(\lambda)-w(\lambda'))$ cancels this singularity, 
so omitting the $i=j$ term in the discrete sum is equivalent to taking the principal-value limit of the integral. 
The matrix $M$ is real and symmetric, and its spectrum converges
to the continuum eigenvalues $\Delta_n$ as the number of quadrature points $Q$ increases.

\begin{figure}[H]
\centering
\includegraphics[width=.6\textwidth]{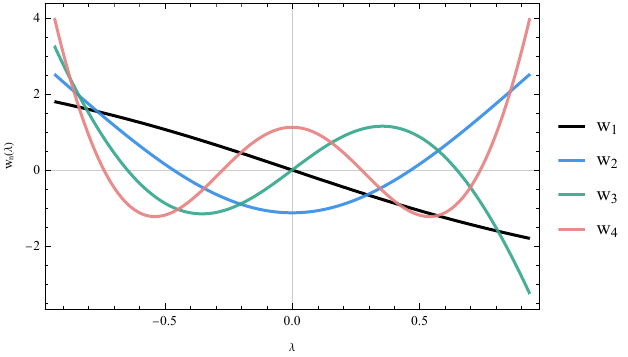}
\caption{Plot of the first four long string eigenvectors numerically computed from Eq.~\eqref{finaleigenvalueequation}.}
\label{longstringeigenvectors}
\end{figure}

The first four eigenvectors computed from $M$ are shown in Figure~\ref{longstringeigenvectors}. The space of eigenvectors is the space of $\tilde{\mathbf{w}}_n=(\tilde{w}_1,\dots, \tilde{w}_Q)$ that are square integrable on the domain $ [-\lambda_{max}, \lambda_{max}]=[\lambda_-(e_f), \lambda_+(e_f)]$. Finally, to compute $h_{abcd}$ (Table~\ref{longStringCouplingCoeffs}), I compute the discrete sum corresponding to the integral in Eq.~\eqref{habcd}.

\begin{table}[H]
\centering
\begin{tabular}{|c|c|c|c|c|c|}
\hline
Quadrature & $\Delta_1$ & $h_{1111}$ & $h_{1212}=h_{2121}$ & $h_{1313}=h_{3131}$ & $h_{2222}$ \\
\hline
900 & 2.1249 & 0.328753 & 0.621461 & 0.592918 & 1.12301 \\
\hline
1000 & 2.12523 & 0.328365 & 0.621169 & 0.59258 & 1.12144 \\
\hline
1200 & 2.12573 & 0.327784 & 0.620731 & 0.592072 & 1.11909 \\
\hline
\end{tabular}
\caption{Long string coupling coefficients and the first energy eigenvalue for coupling $g=2$ numerically computed from Eq.~\eqref{habcd} using the Nyström method (Ref.~\cite{Atkinson1997}) with Gaussian quadrature. \label{longStringCouplingCoeffs}}
\end{table}

From Eq.~\eqref{partitionfunc}, the low temperature expansion of the energy expectation value in the long string effective theory in the planar limit is
\begin{equation}
\frac{E_{L.T.}(\beta)}{N^2} = e_0 + \Delta_1  e^{-\beta \Delta_1} + \mathcal{O}(e^{-2 \beta \Delta_1}, e^{-\beta \Delta_2})
\end{equation}
where $e_0$ is the ground state energy. 
I refer to 
\begin{equation}
\frac{E_{L.T.}^{(1)}}{N^2} \coloneqq e_0 + \Delta_1 e^{-\beta \Delta_1} 
\end{equation}
as the first order approximation and 
\begin{equation}
\frac{E_{L.T.}^{(2)}}{N^2} \coloneqq e_0 + \Delta_1 e^{-\beta \Delta_1} + \Delta_2 e^{-\beta \Delta_2} + [\Delta_1 + h_{1111} (1-2\beta \Delta_1)]e^{-2\beta \Delta_1}
\end{equation}
as the second order. In general I refer to $c(\beta)e^{-a\beta \Delta_b}$ as an $ab$-order term.

\begin{figure}[H]
\centering
\includegraphics[width=.6\textwidth]{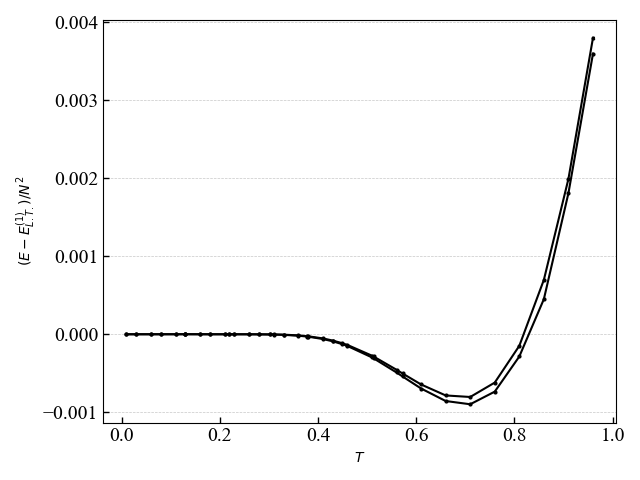}
\caption{The difference between QICS bootstrap bounds from system size $L=12$ and the first order approximation with values in Ref. \cite{Marchesini:1978ud}.}
\label{oneMatrixg2L12firstresiduals}
\end{figure}

\begin{figure}[H]
\centering
\includegraphics[width=.515\textwidth]{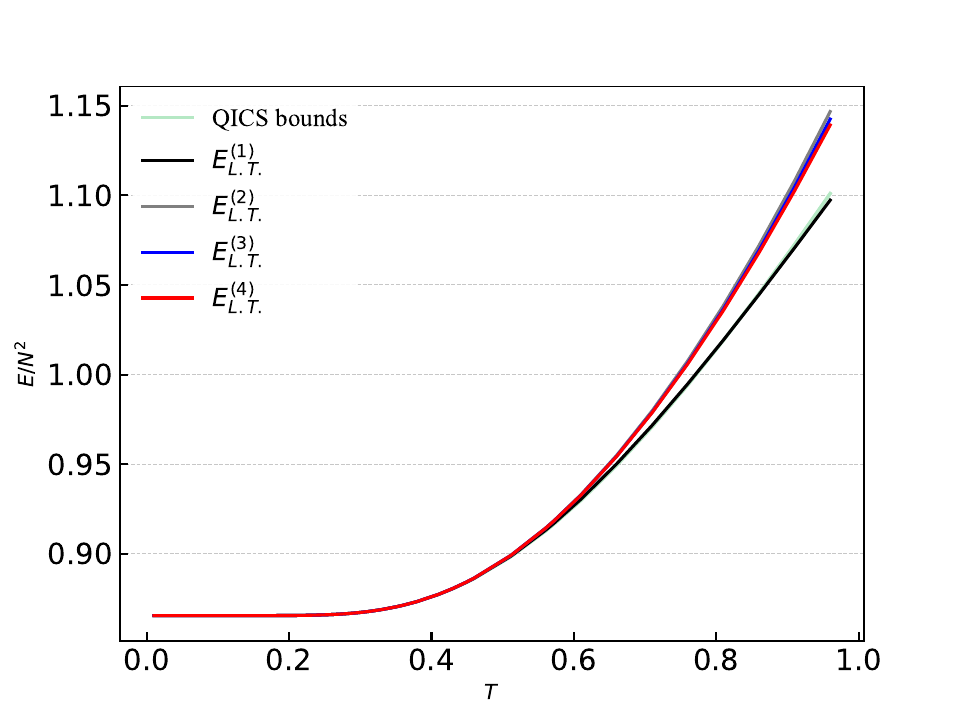}
\includegraphics[width=.475\textwidth]{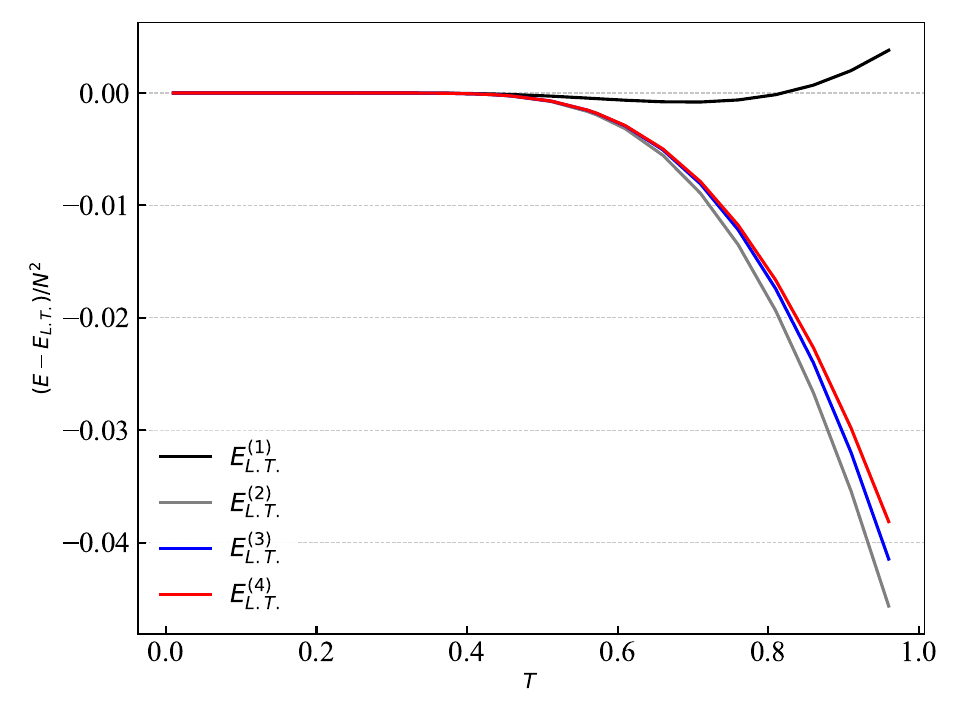}
\caption{QICS bounds compared to analytic approximations of the energy with values from Ref. \cite{Marchesini:1978ud} and long string coupling coefficients from Table~\ref{longStringCouplingCoeffs}.}
\label{oneMatrixg2L12ELT1234}
\end{figure}

Bootstrap bounds for system size $L=12$ found using QICS show good agreement with the first order long string energy approximation $E_{L.T.}^{(1)}$ (Figure~\ref{oneMatrixg2L12firstresiduals}). The second order approximation overshoots the bootstrap bounds, while successive approximations bring $E_{L.T.}$ back down (Figure~\ref{oneMatrixg2L12ELT1234}). Adding the first long string interaction term $[\Delta_1 + h_{1111} (1-2\beta \Delta_1)]e^{-2\beta \Delta_1}$ to the first order approximation improves the fit for low $T$ and allows us to extract values for $e_0$, $\Delta_1$, and $h_{1111}$ consistent with their analytic/numerical values using Powell optimization of a penalized chi-squared cost function on the low $T$ part of the graph with an initial guess of $[e_0,\Delta_1,h_{1111}]=[1,1,1]$ (Tables~\ref{oneMatrixL12FitLT} and \ref{oneMatrixL12FitFull}). While this method of fitting is not particularly sensitive to higher-order terms, these terms could potentially be extracted using semidefinite programming on the spectral function as in Ref. \cite{Collier_2018}.

\begin{table}[H]
\centering
\begin{tabular}{|c|c|c|}
\hline
 & Bootstrap results & Analytic/numerical \\
\hline
$e_0$ & $0.865457750210 \pm 3\times 10^{-7}$ & 0.8654577 \\
\hline
$\Delta_1$ & $2.1283360 \pm 2 \times 10^{-4}$ & 2.1281936 \\
\hline
$h_{1111}$ & $0.32731 \pm 7 \times 10^{-2}$ & 0.3278 \\
\hline
\end{tabular}
\caption{Fit parameters from $L=12$ bootstrap bounds fit to the model $\frac{E}{N^2} \coloneqq e_0 + \Delta_1 e^{-\beta \Delta_1} + [\Delta_1 + h_{1111} (1-2\beta \Delta_1)]e^{-2\beta \Delta_1}$ using Powell optimization of a penalized chi-squared cost function on the low temperature part of the graph. The errors were determined by fitting the upper and lower bounds. The analytic values for $e_0$ and $\Delta_1$ are from Ref. \cite{Marchesini:1978ud}, and the numerical value for $h_{1111}$ was computed using a quadrature of 1200 (Table~\ref{longStringCouplingCoeffs}). \label{oneMatrixL12FitLT}}
\end{table}

\begin{table}[H]
\centering
\begin{tabular}{|c|c|c|}
\hline
 & Bootstrap results & Analytic/numerical \\
\hline
$e_0$ & $0.865457750194 \pm 3 \times 10^{-5}$ & 0.8654577 \\
\hline
$\Delta_1$ & $2.1281758 \pm 5 \times 10^{-3}$ & 2.1281936 \\
\hline
$h_{1111}$ & $0.49983 \pm 4 \times 10^{-2}$ & 0.3278 \\
\hline
\end{tabular}
\caption{Fit parameters from $L=12$ bootstrap bounds fit to the model $\frac{E}{N^2} \coloneqq e_0 + \Delta_1 e^{-\beta \Delta_1} + [\Delta_1 + h_{1111} (1-2\beta \Delta_1)]e^{-2\beta \Delta_1}$ using Powell optimization of a penalized chi-squared cost function on the full graph. The errors were determined by fitting the upper and lower bounds. The analytic values for $e_0$ and $\Delta_1$ are from Ref. \cite{Marchesini:1978ud}, and the numerical value for $h_{1111}$ was computed using a quadrature of 1200 (Table~\ref{longStringCouplingCoeffs}). \label{oneMatrixL12FitFull}}
\end{table}

\subsection{Two-Matrix}

\begin{figure}[H]
    \centering
    \begin{subfigure}{0.45\textwidth}
        \includegraphics[width=\linewidth]{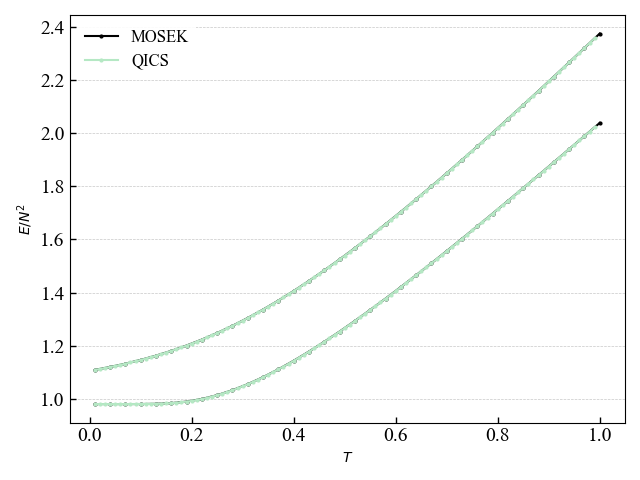}
        \caption{$L=4$}
        \label{twoMatrixq1g01L4}
    \end{subfigure}
    \hfill
    \begin{subfigure}{0.45\textwidth}
        \includegraphics[width=\linewidth]{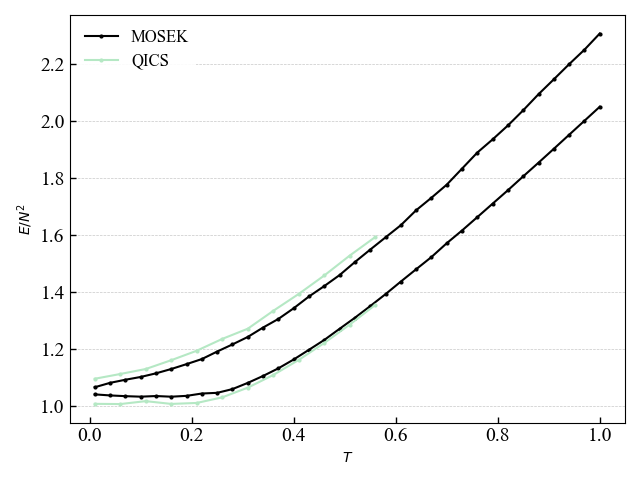}
        \caption{$L=6$}
        \label{twoMatrixq1g01L6}
    \end{subfigure}
    \caption{Thermal energy bounds for $g=0.1$ found using MOSEK with $(3,3)$ logarithmic relaxation and QICS without relaxation for system sizes $L=4,6$. For $L=4$, both solvers converge reliably and produce bounds that agree within numerical tolerance. Imposing the $U(1)$ symmetry constraint allows for bounds to be obtained for $L=6$, but the two-matrix thermal bootstrap still remains numerically challenging at $L=6$.}
    \label{twoMatrixg01L4L6}
\end{figure}

Using the complex operators (Eq.~(\ref{complexops})) instead of the $X_1,X_2,P_1,P_2$ operators and imposing the constraints described in Section~\ref{modelstwomat} significantly reduces the number of final variables from $25$ to $12$ for $L=4$ and $220$ to $81$ for $L=6$. For $L=4$, the bounds found using complex operators agree with the bounds found using $X_1,X_2,P_1,P_2$ operators with the constraints described in Section \ref{modelstwomat}. Furthermore, MOSEK and QICS bounds agree within numerical tolerance (Figure~\ref{twoMatrixq1g01L4}) and can be found in comparable computation time (a couple seconds per optimization). For $L=6$, both MOSEK and QICS fail to converge unless the bounds are found using the complex operators (Eq.~(\ref{complexops})) with $U(1)$ symmetry constraints with a relaxed tolerance. Even then, QICS fails to find bounds past $T=0.56$, and the MOSEK bounds suffer from noise, presumably due to numerical instability (Figure~\ref{twoMatrixq1g01L6}). The $L=6$ bounds took a few minutes per optimization.

\section{Discussion}
A major roadblock in bootstrapping is numerical instability, which can occur when correlator coefficients differ from each other by a large factor. The precise origin of numerical instability is not well understood but often results in solvers struggling to find the feasible region and returning unphysical or no results. Although QICS enforces a stricter version of the thermal (KMS) condition than linear SDP methods and stricter constraints can improve numerical instability, its implementation in double precision ultimately limits the achievable accuracy for larger spaces. While QICS is able to produce rigorous bounds for the one-matrix bootstrap for system size up to at least $L=12$, it encounters numerical instability at $L=6$ in the two-matrix bootstrap. Due to the rapid growth of the operator basis, the two-matrix bootstrap is a substantially more demanding problem and would likely benefit from the use of an arbitrary precision solver. Therefore, a method of solving the bootstrap that balances both precision and nonlinear constraints is still desired. 

Beyond numerical issues, the results demonstrate that nonlinear conic methods can extend thermal bootstrapping beyond the reach of conventional linear relaxations. In the one-matrix case, QICS reproduces previously obtained energy bounds at lower system sizes and enables the computation of new bounds at larger $L$ where conventional linear SDP solvers encounter numerical instability. This indicates that enforcing the KMS constraint directly through a nonlinear conic formulation provides a promising framework for thermal bootstrap calculations at higher truncation levels. For the two-matrix model, imposing additional symmetries, such as $U(1)$ charge conservation, significantly reduces the dimensionality of the constraint space and allows bounds to be obtained at system size $L=6$. These advances represent a necessary step toward extracting higher long string excitations and couplings as well as observables in matrix models dual to theories of black holes. \\

The code used for this paper is available at https://github.com/smadams821/Thermal-Bootstrap-of-Large-N-MQM and archived at Zenodo with DOI 10.5281/zenodo.17497230.

\acknowledgments

I'd like to thank Minjae Cho, Joshua Sandor, and Xi Yin for their guidance and support. I am also grateful to Minjae Cho for allowing me to use his matrix building code and to Xi Yin for suggesting this project. Thank you to Barak Gabai, Henry Lin, Zechuan Zheng, and Kerry He for their helpful comments. Thank you to the Mellon Mays foundation for funding this project.

\bibliographystyle{JHEP}
\bibliography{biblio}
\end{document}